# Mutarotational Kinetics and Glass Transition of Lactose

Ronan LEFORT[1*], Vincent CARON[2**], Jean-François WILLART[2], Marc DESCAMPS[2]

[1]GMCM - UMR CNRS 6626, [2]LDSMM - UMR CNRS 8024 – ERT 1018

*Université de Rennes 1, Campus de Beaulieu, F-35042, Rennes Cedex.

**Université de Lille 1, Cité Scientifique, F-59655, Villeneuve d'Ascq cedex.

## Abstract

We report for the first time real time *in situ* and quantitative measurements of the mutarotation reaction of lactose in the solid state. The experiments have been performed by $^{13}$C NMR. We show that mutarotation is initiated on heating the amorphous state, and reaches chemical equilibrium close above the glass transition temperature $T_g$. We do not observe this transformation when starting from stable crystalline states. The final ratio of $\alpha$ and $\beta$ anomers is 1:1, which suggests that the energy profile of the mutarotation reaction pathway in the solid state is actually different from the mechanism proposed for aqueous solution. This chemical equipartition is reached before the crystallization into the corresponding 1:1 molecular compound. These new data clearly illustrate the interrelation between the chemical molecular properties, the physical state of the material, and the relaxational dynamics of the glass.

# Introduction

Lactose, *4*-O-*β*-D-Galactopyranosyl-*α*-D-glucopyranose, is a natural disaccharide (milk sugar) that presents at the molecular level two isomeric forms ($\alpha$ or $\beta$) differing by the conformation of the anomeric $C'_1$ carbon of the glucopyranose ring. The interconversion from one anomer to the other is called mutarotation, in reference to the optical rotation it induces in aqueous solution. This property results in a very rich variety of crystalline states of lactose, including monohydrate [1], anhydrous $\alpha$ and $\beta$ pure forms [2-4], and several $\alpha/\beta$ defined molecular compounds [2, 3, 5]. One of the simplest mechanisms of mutarotation that was proposed long ago for D-glucose [6, 7], and retained among others for lactose [8] is schematically represented in Figure 1.

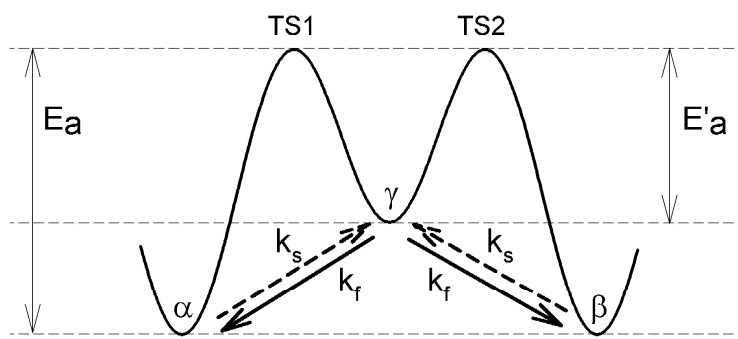

**Figure 1. Energetical pathway for the interconvertion between the $\alpha$ and $\beta$ forms of lactose (mutarotation) via an intermediate free aldehyde form of the glucose residue (g). Two transition states TS1 and TS2 are separated from the other forms by energy barriers $E_a$ and $E'_a$. The rate constants for the intermediate reactions are $k_s$ and $k_f$.**

This reaction pathway assumes the existence of an intermediate $\gamma$ free aldehyde form of the glucose residue, separated from the $\alpha$ and $\beta$ ring forms by two transition states TS1 and TS2. Without losing much generality, the energetic profile defined in Figure 1 is assumed to be symmetric, with only two relevant activation parameters $E_a$ and $E'_a$. A few years ago, values of $E_a \approx 30$ kcal/mol and $E'_a \approx 20$ kcal/mol were evaluated by an *ab initio* numerical approach taking into account partial hydration of lactose isomers and transition states [7]. Recently, it has been reported that mutarotation of lactose could be observed in the solid state, after heating a monohydrate crystalline sample [8, 9]. It was also suggested that it could be induced at room temperature by mechanical stress, during ball-milling amorphization [10], although this last point remains debated [11]. The existence of anomeric interconversion during solid state transformations of lactose deserves particular attention, since it might play an important role in the states diagram, or act as a perturbation on the crystallization and transitions kinetics [12]. Especially, the observation of mutarotation of lactose in the amorphous state is linked to the more general problem of diffusion-controlled reactions near the glass transition of molecular systems. This problem has gained renewed interest, especially in the fields of pharmacy, where it helps understanding the denaturation rates of active species trapped in amorphous excipients [13], or in biophysics, where many studies in the past decades have assessed the remarkable bioprotective skills of many disaccharides [14-16]. Indeed, much interest was focused on the specific dynamical properties of proteins embedded in a carbohydrate highly viscous or glassy matrix [17-20]. Those studies revealed that internal chemical reactions that take place on proper sites of the protein were submitted to highly dispersive kinetics [21, 22], that could be empirically described far below $T_g$ by large distributions of activation barriers.

This article reports for the first time a direct experimental evidence of mutarotation of lactose in the solid state. We present in the first section nuclear magnetic resonance results and discuss the conditions that make the mutarotation reaction possible in the glass. In the second section, we discuss the different temperature regimes of the mutarotation kinetics in the amorphous state, and show how this chemical property of the lactose molecule is intimately interrelated to the molecular mobility and is thus strongly affected by passing through the glass transition domain.

## Materials and Methods

**Crystalline α lactose monohydrate (αLM)** was purchased from Sigma and used without further treatment.

**Crystalline anhydrous α lactose (αL)** was obtained by dehydration of crystalline αLM. We performed dehydration by blowing dry gaseous methanol through 20 g of α-lactose monohydrate during 3 hours at 64.7°C. The remaining traces of methanol were then removed by placing the sample under vacuum ($10^{-3}$ mm Hg) at 20°C during 12 hours. Thermogravimetric measurements have indicated that the lactose thus obtained was free of water and methanol.

**Amorphous α lactose (αLA)** was obtained by ball milling of αL during 30 h as described in reference [11].

**The ball milling** was performed in a high energy planetary mill (Pulverisette 7 – Fritsch) at room temperature (RT) under dry nitrogen atmosphere. We used $ZrO_2$ milling jars of 43 cm$^3$ with seven balls (Ø=15 mm) of the same material. 1 g of material was placed in the planetary mill corresponding to a ball / sample weight ratio of 75:1. The rotation speed of the solar disk was set to 400 rpm which corresponds to an average acceleration of the milling balls of 5 g.

**The $^{13}$C CPMAS spectra** were recorded on a Bruker AV400 solid-state NMR spectrometer operating at 100.61 MHz for $^{13}$C, using a 4 mm CPMAS probe. Before each experiment, the $ZrO_2$ MAS rotor was dried in oven at 120°C. All samples were spun at the magic angle at a 5 kHz rotor speed. A 1 ms cross-polarization contact pulse with power ramp, and a relaxation delay of 50 s were used. Dipolar decoupling of the protons during the acquisition was achieved with a high power multipulse tppm train. No apodization of FID signals was applied prior to Fourier Transform.

**The spectral deconvolution** of the $^{13}$C CPMAS spectra was achieved by means of gaussian approximation of the actual NMR lineshape, and integration of the best model refinements was realized by means of the DMFIT software package [23].

## Results

The Figure 2 shows the $^{13}$C CPMAS spectra of crystalline αL and αLM [2]. The lowest field NMR line (105-107 ppm) is attributed to the galactose $C_1$ carbon. The glucose $C'_4$ NMR signal appears in the 80-90 ppm range, as a doublet for αL (two inequivalent lactose molecules/cell [3]), and a singlet for αLM. The anomeric $C'_1$ carbon of the glucose residue resonates between 92 and 98 ppm. It is known [2] that this line for the *β* anomer is significantly shifted downfield compared to the *α* anomer. It is clear on Figure 2 that only the resonance of *α*-lactose is evidenced from the NMR results, insuring the anomeric purity of the initial studied samples. The resonances observed between 55 and 65 ppm are assigned to the $C_6/C'_6$ carbons (primary alcohol groups), and between 65 and 78 ppm to the remaining pyranose rings carbons. The $^{13}$C CPMAS spectra show characteristic differences for αL and αLM, and can be used as reference spectra to identify those two crystalline forms.

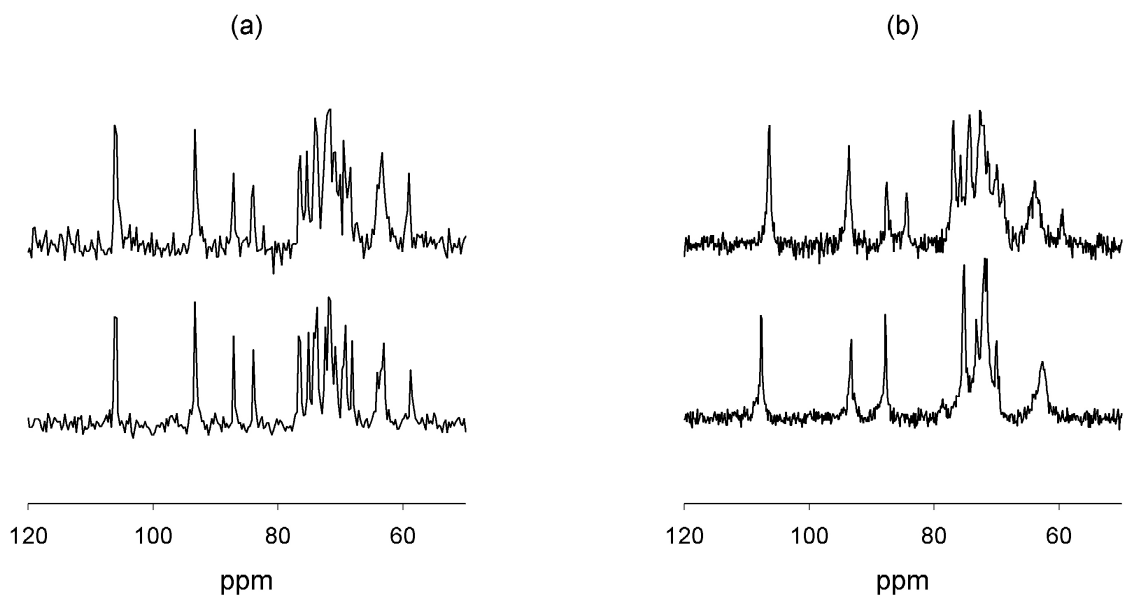

**Figure 2.** $^{13}$**C CPMAS spectra of crystalline forms of** α**-lactose. (a) : stable anhydrous** α**-lactose at room temperature (lower line), then heated to 160°C (upper line) ; (b) :** α**-lactose monohydrate at room temperature (lower line), then transformed into stable anhydrous form when heated up to 140°C (upper line)**

On heating αL, no change in the CPMAS spectrum is noticed until the fusion (and degradation) of the sample is observed. On heating αLM, dehydration takes place between 120 and 140°C. It is clear from Figure 2(b) that this dehydration results in a complete transformation from αLM into αL. As a consequence, the obtained αL remains stable until its fusion premices (around 170°C). As an evidence, no mutarotation from $\alpha$ to $\beta$ anomeric form of lactose is observed upon heating the crystalline forms αL and αLM.

The situation is markedly contrasted when the sample is initially amorphous. We showed elsewhere [11] that the quasi pure α-anomeric form of lactose in an amorphous state can be obtained by ball-milling under dry nitrogen atmosphere. Particular care was paid in the present work to exclude mutarotation of lactose due to partial solution in residual liquid water. The NMR rotors were therefore dried in oven before the experiments, and the milled powder was also dried at 80°C on a kofler bench before filling in the NMR rotor, in order to prevent accidental humidity adsorption.

Figure 3(a) shows the $^{13}$C CPMAS spectra of amorphous lactose following slow heating up to 132°C (about $T_g$+20 K). In comparison with Figure 2, the NMR response at room temperature displays broadened lines, reflecting the underlying distribution of conformational degrees of freedom (especially around the glycosidic bond [24-26]). The NMR lines attributed to galactose $C_1$ and glucose $C'_1$ ($\alpha$ or $\beta$) carbons remain however sufficiently resolved to be quantitatively exploited. On heating, several changes are noteworthy: first, the resonance at 98 ppm characteristic of $\beta$-lactose molecules grows with temperature. This constitutes the first direct observation that mutarotation of lactose actually occurs in the amorphous solid state. As local structural environments and dynamical fluctuations of local dipolar field are likely to be very similar for $C'_1$ carbons in $\alpha$ or $\beta$ conformations, we take the cross-polarization transfer between protons and $^{13}$C nuclei to be identical. In this case, the integrated NMR signal is linearly related to the number of $^{13}$C nuclei, and model integration of the $\alpha$ (93 ppm) and $\beta$ (98 ppm) lines provides a quantitative analysis of the $\beta$-lactose molecular fraction in the sample, and so the progress of the mutarotation reaction in the solid state. Second, at 132°C (i.e. above $T_g$), the CPMAS spectrum becomes more structured, especially in the pyranose ring carbons region (65-78 ppm). As illustrated in Figure 3(b), this thinning of the NMR lines vanishes on cooling the sample from 132°C down to room temperature while the the peak characteristic of the beta anomer (at 98 ppm) persists. The corresponding spectrum is thus characteristic of amorphous lactose which however contains a much higher β-lactose fraction than the amorphous lactose obtained just after milling. This indicates that the mutarotation observed upon heating is not reversible upon cooling. It is also the proof that the observed thinning of the lines at high temperatures has a dynamical origin and must be attributed to the structural relaxations which becomes increasingly faster above Tg and thus begin to average the chemical shift anisotropy tensor.

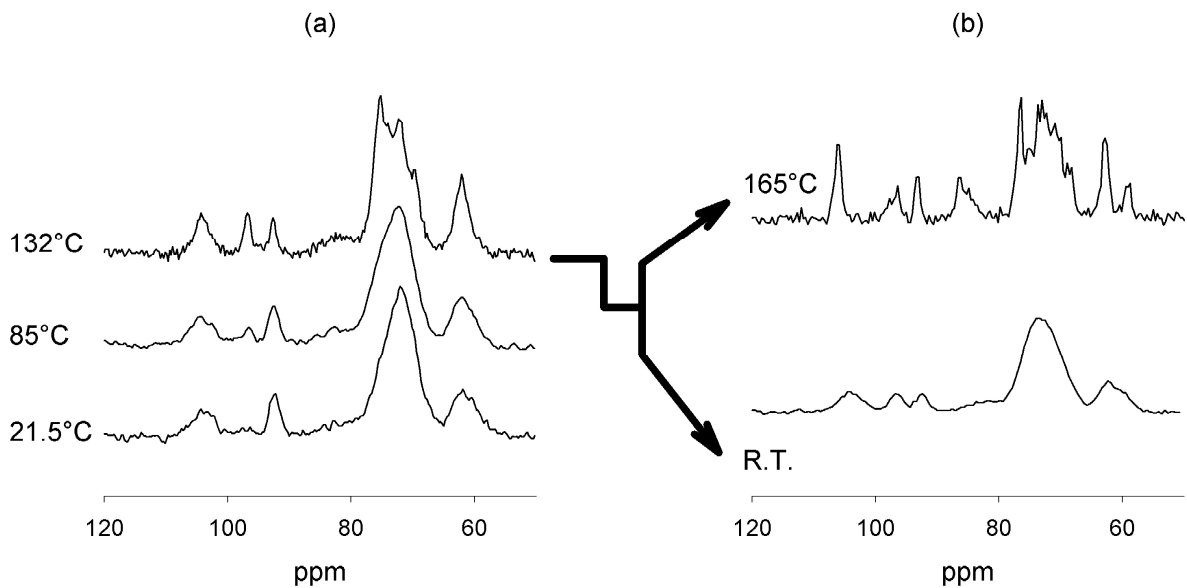

**Figure 3. (a) Evolution on temperature of the $^{13}$C CPMAS Spectra amorphous α-lactose obtained by milling. The NMR line growing around 96 ppm is assigned to β-lactose molecules. After heating up to 132°C, the sample is either quenched down to room temperature, of heated up to 165°C, where crystallization into the**

α:β=1:1 molecular compound is observed. The corresponding $^{13}$C CPMAS spectra after returning to RT are presented in (b).

On the other hand, if the heating is pursued over 165°C, crystallization occurs as evidenced by differential scanning calorimetry [11]. The NMR spectrum of the recrystallized sample (Figure 3(b)) shows that the α:β=1:1 ratio is reached, indicating the formation of a molecular compound of defined stoichiometry. The crystallographic structure of this α:β=1:1 molecular compound was solved recently by means of powder X-ray diffraction [3]. Figure 4 compiles the evolution of the β-lactose fraction measured upon heating in the amorphous state by solid state NMR. This fraction increases with temperature, and reaches the value of 50% just above Tg at around 132°C (in the amorphous state). This value remains unchanged on further heating, or when cooling back the sample before it recrystallizes.

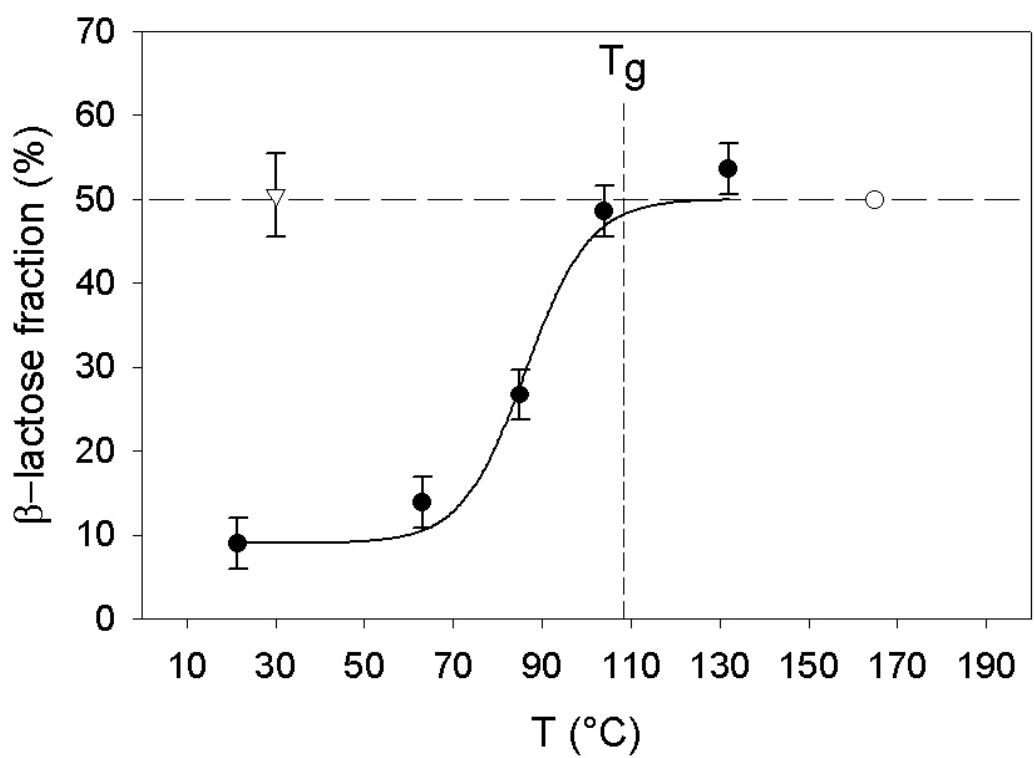

**Figure 4.** Evolution with temperature of the β-lactose fraction in the amorphous state, as measured by NMR (filled circles). After reaching 132°C, the 1:1=α/β amorphous sample is either quenched to room temperature (down triangle), or further heated so that it transforms toward the α:β=1:1 crystalline phase (open circle). The solid line is a guide for the eye.

## Discussion

The NMR data presented here show that the amorphous α-lactose undergoes a strong mutarotation effect upon heating while this effect does not occur in the corresponding crystalline forms (αL and αLM). In the amorphous state the mutarotation process toward the $\alpha:\beta=1:1$ ratio is fully completed before the onset of recrystallization toward the defined $\alpha:\beta=1:1$ molecular compound. This suggests that the mutarotation of lactose in the solid state does not imply a cooperative reaction pathway governed by a many-body mechanism nor long range interactions. In aqueous solution, the mutarotation kinetics can be followed by several techniques like optical rotation or NMR [27, 28]. The equilibrium is reached in a few seconds, and the $x_{eq} = \left(\dfrac{\beta}{\alpha}\right)_{eq}$ ratio of equilibrium populations of lactose anomers is temperature dependent and greater than 1 [8], which suggest a different hydration shell for both forms of the molecule. In the dry solid state, water cannot be evoked as a reaction intermediate. One can distinguish two different regimes corresponding to two different temperature ranges: first, from room temperature to about 120°C, the measured β-lactose fraction is found temperature dependent. Lerk and co-workers also found that it was highly dependent on the preparation scheme of the sample (initial $\beta$-lactose content) [27]. Second, above 120°C, the population ratio of the two forms is found independent on both the temperature and the initial state. These observations suggest that the equilibrium populations $\alpha$:$\beta$=1:1 are reached at sufficiently high temperatures, while those reported below 120°C on Figure 4 are out of equilibrium on the time scale of the NMR experiment. In order to assess this out of equilibrium character, isothermal measurements at 85°C were performed during several hours, starting from a sample containing about 35% of β-lactose when this temperature was reached. Figure 5 shows that the mutarotation reaction indeed advances with a rate of about 3% per hour, toward an equilibrium value consistent with that found at higher temperatures (50%).

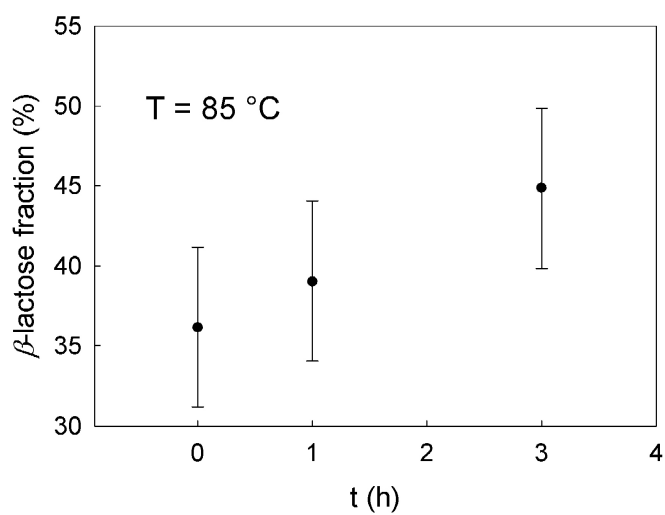

**Figure 5 : Isothermal evolution of the β-lactose content measured at 85°C on a ball-milled amorphous sample.**

A simple way to account for these experimental evidences is to state that mutarotation of lactose, although being a local chemical process, strongly couples to the structural relaxation of the amorphous state. As a matter of fact, it is clear from Figure 4 that $T_g$ represents a crossover temperature for the mutarotational kinetics in solid state lactose. As a consequence, pure amorphous α-lactose can only exists in non-equilibrium conditions, and requires to be obtained directly far below $T_g$ where the $\alpha \leftrightarrow \beta$ reaction is kinetically strongly hindered. That is why, to our knowledge, milling is actually the only amorphization process able to produce pure amorphous α-lactose. The usual routes (thermal quench of the liquid, freeze or spray drying …) always require either to heat the sample, either to put it in solution which produces unavoidably a large amount of β-lactose. This is also a clue for rationalizing the importance of humidity in the amorphization process of lactose. Indeed, the key role played by residual water on facilitating mutarotation in the solid state can now rather be understood in terms of a plasticizing effect (decrease of Tg) than as a direct chemical intermediate. This supports prior evidence that a thorough control of the amorphization process of lactose can only be reached in a dry atmosphere [11].

The mutarotation of lactose in the solid-state represents to our knowledge the first example of reaction kinetics in a molecular carbohydrate glass, that presents such crossover from a low temperature regime where the reaction is controlled by molecular diffusion, towards a high temperature regime where is it classically controlled by a mass action law (for a review, see e.g. [22]). Around $T_g$ and above, such coupling of the reaction activation pathway to the solvent structural relaxation has often been described in the high viscosity ($\eta$) limit using the Kramers model by $k \sim \eta^{-1}$ [29]. Sumi and Marcus [30, 31] proposed another form for the effective rate constant $k_{eff}$ in viscous solvents, very similar to the Rabinowitch equation [32] widely employed in polymer science to describe curing reactions :

$$\frac{1}{k_{eff}} = \frac{1}{k_m} + \frac{1}{k_d} \qquad (1)$$

The Sumi-Marcus model accounts for a crossover from a high temperature regime where the reaction is mass controlled and where the rate constant $k_m$ can be derived from standard kinetic models like transition state theory (TST) [33], to a low temperature regime, where the reaction becomes diffusion controlled, and the specific rate constant $k_d$ couples to the viscosity through a power law $k_d \sim \eta^{-\alpha}$. This last expression unifies the two aspects mentioned above, namely the coupling to the structural relaxation in the undercooled liquid approaching the glass transition, and the global account through the exponent $0 \leq \alpha \leq 1$ for the distribution of the energy barriers due to local fluctuations of the part of intramolecular or conformational degrees of freedom that affect the reaction.

In pure α-lactose, the reactant species do not consist in diluted invited solutes. The simple assumption that $\alpha$ and $\beta$ states are isoenergetic in the model presented in Figure 1 accounts alone for the experimental evidence that the equilibrium interconversion percentage is 50%. On the time scale of the NMR experiment (several hours), it is clear that only the long time evolution of the reaction is probed. While equation (1) can provide a reasonable phenomenological description of the different regimes of the reaction, a more quantitative analysis on this time scale is difficult. For temperatures far below $T_g$, other factors could also have some influence on the reaction kinetics, like secondary relaxational processes.

## Conclusion

We reported in this paper the first *in situ* observation of mutarotation of lactose in the solid state. This mutarotation is found to occur in the amorphous states but not in the corresponding crystalline phases showing the strong influence of the physical state on the mutarotation process. The equilibrium ratio between anomeric $\alpha$ and $\beta$ forms of lactose is found to be 1:1 in the amorphous state before crystallization occurs into the corresponding defined molecular compound. This 1:1 anomeric ratio is therefore not stabilized during crystallization by long range cooperative effects, but rather demonstrates the local energetic equivalence of the two forms in the solid at all temperatures. This result suggests that the reaction energetic profile in solid amorphous lactose is different from mutarotation in aqueous solution, where the $\beta$ form is slightly preferred. A quantitative measurement of the $\beta$-lactose content could be assessed by NMR and followed on a significant temperature range. Two regimes could be evidenced : one at low temperature where the mutarotation reaction remains far from thermal equilibrium on the hour time scale and is controlled by the molecular diffusion, and a second one above $T_g$ where the reaction is fast, and classically controlled by a mass action law. We showed that our data qualitatively agree with the general results established in the frame of dispersive kinetics models, accounting for a coupling of the reaction rates to the structural relaxation near the glass transition. This study suggests that chemical reactions could appear in certain cases as a very powerful kinetic probe of the molecular relaxation local mechanisms involved in the glass transition process.

## Acknowledgements

All experiments analyzed in the present report were carried out on the equipments of the CCRMN (common NMR center) of the University of Lille 1. We are thankful to B. Revel for precious technical support.